\documentclass[a4paper,aps,pra,showpacs,twocolumn,floatfix,superscriptaddress]{revtex4}

\usepackage[dvips]{graphicx}

\begin{document}

\title{
Optically guided atom interferometer tuned to magic wavelength
}

\author{Tomoya Akatsuka}
\altaffiliation{Present address: NTT Basic Research Laboratories, Nippon Telegraph and Telephone Corporation, 3-1 Morinosato Wakamiya, Atsugi, Kanagawa 243-0198, Japan}
\affiliation{Department of Applied Physics, Graduate School of Engineering, The University of Tokyo, Bunkyo-ku, Tokyo 113-8656, Japan}
\affiliation{Quantum Metrology Laboratory, RIKEN, Wako, Saitama 351-0198, Japan}

\author{Tadahiro Takahashi}
\affiliation{Department of Applied Physics, Graduate School of Engineering, The University of Tokyo, Bunkyo-ku, Tokyo 113-8656, Japan}
\affiliation{Quantum Metrology Laboratory, RIKEN, Wako, Saitama 351-0198, Japan}

\author{Hidetoshi Katori}
\altaffiliation{E-mail: katori@amo.t.u-tokyo.ac.jp}\affiliation{Department of Applied Physics, Graduate School of Engineering, The University of Tokyo, Bunkyo-ku, Tokyo 113-8656, Japan}
\affiliation{Quantum Metrology Laboratory, RIKEN, Wako, Saitama 351-0198, Japan}

\begin{abstract}
We demonstrate an atom interferometer operating on the $^1\!S_0-{}^3\!P_0$ clock transition of $^{87}$Sr atoms in a ``magic'' optical guide, where the light shift perturbations of the guiding potential are canceled.
As a proof-of-principle demonstration, a Mach-Zehnder interferometer is set horizontally to map the acceleration introduced by the focused optical guide. 
This magic guide interferometer on the clock transition is applicable to atomic elements where magic wavelengths can be found.
Possible applications of the magic guide interferometer, including a hollow-core fiber interferometer and gradiometer, are discussed.
\end{abstract}

\maketitle

Technologies supporting miniaturized atomic sensors, such as atomic clocks~\cite{Kna05,Oka14,Szm15}, magnetometers~\cite{Sha07}, and interferometers~\cite{Wan05,McD13,Mar15,Abe16}, are of increasing concern as foundations for future sensing devices. 
A major challenge common to these endeavors is to establish containers for atoms that minimally perturb the measurement outcome~\cite{Oka14,Szm15,Wan05,McD13,Mar15}.
These efforts allow the devices to achieve precision competitive with conventional free-fall apparatuses and access new physics and applications that are not otherwise probed, such as measuring the local magnetic field~\cite{Sha07} and $\mu$m scale gravity~\cite{Pol11,Mar15}.

Atom interferometers (AIs)~\cite{Cro09} consist of the coherent splitting and recombination of atoms followed by atomic state detection.
A photon recoil momentum $\hbar {\bf k}$, with $\hbar$ the as Planck constant and ${\bf k}$ as the wavevector, is conveniently used to establish a Mach-Zehnder interferometer (MZI) for the translational degree of freedom.
Depending on the detection of the output state, AIs are categorized into two types: internal-state or momentum-state labeling. 
The Ramsey-Bord\'{e}~\cite{Bor89,Rie91} and Kasevich-Chu~\cite{Kas91} interferometers project the results onto the two long-lived electronic states $|1\rangle$ and $|2\rangle$ of atoms, while the Bragg interferometer~\cite{Gil95,Tor00,Chi11,Maz15,Est15} projects onto the atomic momentum states ${\bf p}$ and ${\bf p}+\hbar {\bf k}$.

Owing to the ultracold atoms with a very small momentum spread $|\Delta {\bf p}|\ll |\hbar {\bf k}|$ produced by the Bose-Einstein condensation and delta-kick cooling~\cite{Kov15}, the state-of-the-art MZIs~\cite{Hos16,Abe16} rely on the Bragg scheme combined with a large momentum- transfer technique~\cite{Dav93,Gil95,McG00,Mul08,Chi11,Est15}. 
As long as the spatial imaging of atoms resolves the two momentum states, the Bragg scheme using the same electronic states is more advantageous than the electronic state labeling, as the use of the same electronic state makes the interferometer insensitive to the electromagnetic (EM) perturbations.
Taking this advantage, magnetically~\cite{Wan05} or optically~\cite{McD13,Mar15} guided interferometers, which offer an extended evolution time with a small interaction volume, apply the Bragg scheme to be free from the EM perturbation of the guiding fields, such as the Zeeman shift and AC Stark shift.

Once an electronic-state independent guide is realized, the internal state labeling~\cite{Bor89,Kas91} is particularly beneficial for miniaturizing AIs, as it does not require the spatial imaging of atoms, which necessitates a large volume, to discriminate the momentum states.
In this letter, we demonstrate an optically guided MZI on the $^1\!S_0-{}^3\!P_0$ clock transition of $^{87}$Sr atoms by operating the optical guide at the magic wavelength.
As the AC Stark shift of the guide becomes equal for the two states~\cite{Kat03,Kat11} as illustrated in Fig.~1, the perturbation of the optical guide does not introduce an extra phase shift in the interferometer.
As a proof-of-principle demonstration, we measured the position-dependent acceleration of atoms in the focused ``magic guide'' [see Fig.~1(a)].
We discuss prospects for the ``magic fiber-guide'' as an alternative design for future miniaturized interferometers [see Fig.~1(b)].
In particular, we mention application to a gradiometer to reject the phase noise of a clock laser that fundamentally limits the performance of AIs using a single-photon clock transition.

\begin{figure}[t]
\begin{center}
\includegraphics[width=\linewidth]{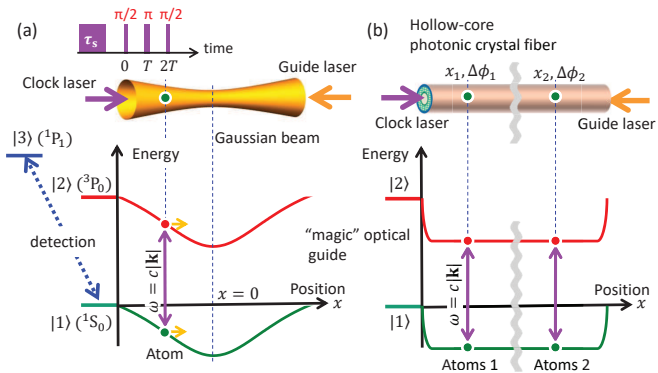}
\caption{
Schematic of an AI with a ``magic'' optical guide that provides the same AC Stark shift on the long-lived states $|1\rangle$ and $|2\rangle$, which correspond to the $^1\!S_0$ and $^3\!P_0$ clock states of Sr and the other alkaline earth (like) atoms. 
Pulsed excitation of the clock states with a $\pi/2-\pi-\pi/2$ sequence realizes a Mach-Zehnder interferometer. 
Output of the interferometer is mapped onto the clock states, which is detected by driving the $|1\rangle-|3\rangle$ transition.
(a) In a free-space optical guide, a finite Rayleigh range introduces axial acceleration (yellow arrows) on atoms.
(b) When a hollow-core photonic crystal fiber is applied, atoms mainly sense external acceleration because of the negligible attenuation of the guiding potential. 
Such a magic-guide interferometer is best applied for a gradiometer consisting of two atomic samples, i.e., atoms 1 and 2, where the clock-laser instability is common-mode rejected for the measurement outcome $\Delta a=a_2-a_1$.
}
\end{center}
\end{figure}%

Figure 2 shows a schematic of the experiment. 
We set an optical guide perpendicular to gravity and operate the guide at the magic wavelength $\lambda_{\rm m}=813.4$~nm \cite{Kat11} of the $^1\!S_0 (F=9/2)-{}^3\!P_0 (F=9/2)$ clock transition of $^{87}$Sr atoms.
The intensity of this guide laser is given as $I_{\rm{g}}(x,y,z)=2P_{\rm{g}}/[\pi w^2(x)] \exp[-2(y^2+z^2)/w^2(x)]$, with a waist diameter of $2w_0=70$~$\mu$m, a beam radius of $w(x)=w_0[1+(x/x_{\rm R})^2]^{1/2}$, a Rayleigh range of $x_{\rm R}=\pi w_0^2/\lambda_{\rm{m}}=4.7$~mm, and power of $P_{\rm g}$.
The potential energy is given by the sum of the AC Stark shift and the gravitational energy $U(x,y,z)=-\alpha I_{\rm{g}}(x,y,z)/(2\epsilon_0 c)+mgz$, where $\alpha=4.7\times 10^{-39}$~J/(V/m)$^2$~\cite{Saf13} is the electric dipole polarizability, $\epsilon_0$ is the vacuum permittivity, $c$ is the speed of light, $m$ is the atomic mass, and $g$ is the gravitational acceleration.
A guide-laser power of $P_{\rm g}=400$~mW yields the maximum AC Stark shift of 13~$\mu$K. 
Owing to gravity, atoms are trapped around $z=-2.4~\mu$m with an effective potential depth of 8~$\mu$K and a radial oscillation frequency of $\nu_r=320$~Hz at $x=0$.
The axial oscillation frequency is $\nu_x=1.7$~Hz near the beam axis and decreases as the radial motion of atoms increases.

\begin{figure}[t]
\begin{center}
\includegraphics[width=\linewidth]{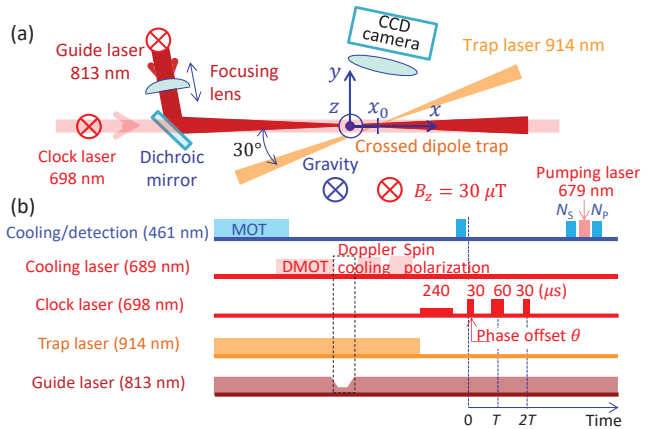}
\caption{
(a) Experimental setup.
A state-insensitive optical guide is set along $x$-axis by a guide laser at 813~nm.
The focal position of the guide laser, which defines $x=0$, is varied by translating a focusing lens.
By adding a trap laser at 914~nm, a crossed dipole trap is formed to prepare spatially localized atoms. 
This crossed dipole trap lies in the $x$-$y$ plane, which is perpendicular to gravity.
A clock laser at 698~nm is introduced along the guide laser.
Sequential pulses start at $t=0$ and excite atoms at $x_0$.
Atomic fluorescence is observed by a CCD camera by driving the $^1\!S_0-{}^1\!P_1$ transition.
(b) Timing chart of the optically guided MZI. 
The widths of the boxes are not drawn to scale.
A phase offset $\theta$ is added to the first clock pulse.
}
\end{center}
\end{figure}%

A clock laser tuned to the $^1\!S_0-{}^3\!P_0$ transition at $\lambda_{\rm{c}}=698$~nm is introduced along the $x$-axis with a beam diameter of 160~$\mu$m.
A $\pi$-pulse with a duration of $\tau_0=60~\mu$s is obtained for a clock laser power of 7~mW.
After applying a velocity-selective $\pi$ pulse with a duration of $\tau_{\rm s}(>\tau_0)$ [see Fig.~1(a)], we apply sequential clock pulses $\pi/2$, $\pi$, and $\pi/2$ with intervals of $T$ and a Rabi frequency of $\Omega_0=\pi/\tau_0$ to realize the MZI.
The interference signal is observed by the population distribution in the clock states, which determines the phase shift~\cite{Pet01} $\Delta\phi=k a T^2$ with $k=2\pi/\lambda_{\rm c}$ and $a$, the acceleration of atoms.
The clock laser is stabilized to a 40-cm-long cavity and has a frequency instability of $\sigma_y=3\times10^{-16}$ at an averaging time of 1~s~\cite{Ush15}.
The pulse duration, power, frequency, and phase of the clock laser are controlled by an acousto-optic modulator.

As the optical guide introduces a position-dependent acceleration $a(x)\approx - \partial U(x,y,z)/\partial x/m$, the phase shift $\Delta \phi$ averages out as the atom cloud spreads along the $x$ axis.
To prepare spatially localized atoms in the optical guide, we introduce a trap laser at 914~nm that intersects with the guide laser with an angle of $30^\circ$, as depicted in Fig.~2(a).
This 350-mW trap laser is focused to a beam diameter of 70~$\mu$m.
The initial atom position $x_0$ with respect to the focal spot of the guide laser, which is taken as $x=0$, is varied by translating the focusing lens of the guide laser.

Figure~2(b) shows a timing chart for the experiment.
$^{87}$Sr atoms are laser-cooled to a few mK in a magneto-optical trap (MOT) on the $^1\!S_0(F=9/2)-{}^1\!P_1(F=11/2)$ transition at 461~nm. 
More than $10^5$ atoms are transferred to a dynamic MOT (DMOT)~\cite{Muk03} on the narrow transition $^1\!S_0(F=9/2)-{}^3\!P_1(F=9/2,11/2)$ at 689~nm.
A few $\mu$K atoms are loaded into the crossed dipole trap consisting of the guide laser and the trap laser.
After the DMOT is turned off, the power of the guide laser is adiabatically decreased to 200~mW in 10~ms and is kept constant for 20~ms to remove atoms that are outside the crossed region, after which the power is recovered to the original value in 10~ms.
For the atoms trapped in the crossed dipole trap, we apply one-dimensional Doppler cooling on the $^1\!S_0(F=9/2)-{}^3\!P_1(F=9/2)$ transition for 80~ms. 
We thus reduce atom temperature to $T_x=0.9(1)~\mu$K along the $x$-axis, which is measured by the width of the atom cloud, as shown in Fig.~3.

We apply a uniform magnetic field of $B_z=30~\mu$T in the $z$ direction to define the quantization axis and apply a $\sigma^+$ circularly polarized laser tuned to the $^1\!S_0(F=9/2)-{}^3\!P_1(F=9/2)$ transition for 80~ms to optically pump the atoms to the $m_F=9/2$ substate in the $^1\!S_0(F=9/2)$ state. 
This laser is frequency-modulated by 10~kHz with a deviation of 100~kHz to cover the Zeeman splitting.
Typically, after the trap laser is turned off, 5,000 atoms remain in the optical guide.

We observe the position of atoms using laser-induced fluorescence (LIF) by driving the $^1\!S_0-{}^1\!P_1$ transition at 461~nm.
Figure~3(a) shows charge-coupled device (CCD) camera images of atoms in the optical guide after the trap laser is turned off at 0~ms.
They relocalize at the opposite side of the potential at 350~ms and return to the initial position at 700~ms.
Figure~3(b) summarizes the position (blue squares) and the width (green circles) of the atoms.
The imperfect relocalization of atoms is due to the inhomogeneity of the axial oscillation frequency $\nu_x$ that depends on the radial position of atoms.
The red line determines the center of the axial oscillation and the frequency $\nu_x=1.4$~Hz that infers the radial temperature of $T_{\rm r}=1.2~\mu$K, as discussed later.

\begin{figure}[t]
\begin{center}
\includegraphics[width=1.0\linewidth]{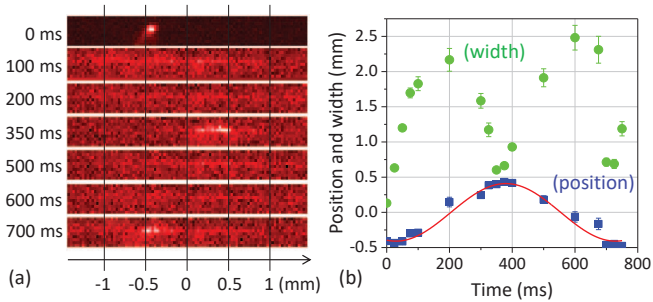}
\caption{
(a) Images of atoms oscillating along the optical guide after the trap laser is turned off.
(b) Position (blue squares) and width (green circles) of atoms at the full width at half maximum are derived from the images.
The solid red line determines the oscillation frequency of $\nu_x=1.4$ Hz and the center of the oscillation.
}
\end{center}
\end{figure}%

Prior to starting the interferometry sequence, we apply a 240-$\mu$s-long $\pi$ pulse to select the velocity group that contributes to the measurement.
Approximately 10\% of atoms with a Doppler width of 3.3~kHz are velocity-selectively excited to the $^3P_0$ state, while the atoms remaining in the ground state are blown away by applying the heating laser at 461~nm [see Fig.~2(b)].
The velocity width in the $x$ direction is $\Delta v_x=2.3$~mm/s, corresponding to an atomic temperature of $T_x=10$~nK.
The initial velocity of the atoms is $v_x(0)=\hbar k/m=6.5$ mm/s.

We operate the MZI by applying three clock pulses with intervals $T$, which start at $t=0$ for atoms at $x(0)=x_0$. 
The outcome of the interferometer is mapped onto the population of atoms $N_S$ in the $^1\!S_0$ state, which is measured by the LIF on the $^1\!S_0-{}^1\!P_1$ transition.
Subsequently, the atoms in the $^3\!P_0$ state are transferred to the $^1\!S_0$ state by exciting the $^3\!P_0-{}^3\!S_1$ transition at 679~nm to measure the number $N_P$ of atoms in the $^3\!P_0$ state.
By giving a phase offset $\theta$ for the first $\pi/2$ pulse with respect to the following two pulses, we measured the fraction of excited atoms as $\kappa(\theta)=N_P/(N_S+N_P)$.

Figure~4(a) shows typical interference fringes for atoms at $x_0= 1.0$~mm in the optical guide with the pulse intervals of $T=0.2$~ms (blue squares) and 1.4~ms (red circles).
They are fitted by $\kappa(\theta)=1-B+A\cos (\theta+ \Delta\phi)$ (solid curves). 
The pulse interval $T$-dependent phase shift $\Delta\phi$ and visibility $A/B$ are measured for several atom positions $x_0$, as shown in Figs.~4(b) and 4(c), respectively.
For reference, the visibility of the MZI without applying the velocity selection is indicated by the empty circles in Fig.~4(c), which shows the improvement of the visibility due to selecting a narrow velocity group. 
Still, the visibility decreases as the pulse interval $T$ increases and nearly fades out for $2T\sim 5$~ms. 
We speculate that this is partly because of the radial motion of atoms, which introduces inhomogeneity to the axial acceleration.
In this regard, loading atoms into the vibrational ground state in the radial motion by applying Bose-Einstein condensates~\cite{Gat09,Dal10} or Raman cooling for the radial motion will improve the visibility.
Additionally, mechanical instability between the atoms and the clock laser causes the loss of contrast by introducing the phase noise owing to the Doppler effect.
Ultimately, the phase noise $\delta \phi\approx 2\sigma_y\omega T$ of the clock laser with its instability $\sigma_y$ and frequency $\omega=c k$ causes the uncertainties for the phase shift and the acceleration $ \delta a\approx \delta \phi/(kT^2)=2\sigma_yc/T$, which has been recognized as the major obstacle for the AI on the single-photon transition~\cite{Rie91} compared with two-photon schemes~\cite{Kas91}.
The relevant uncertainty is estimated to be $\delta a\sim 4\times 10^{-5}~{\rm m/s^2}$ and is significantly smaller than the other experimental uncertainties for the present case. 

\begin{figure*}[t]
\begin{center}
\includegraphics[width=\linewidth]{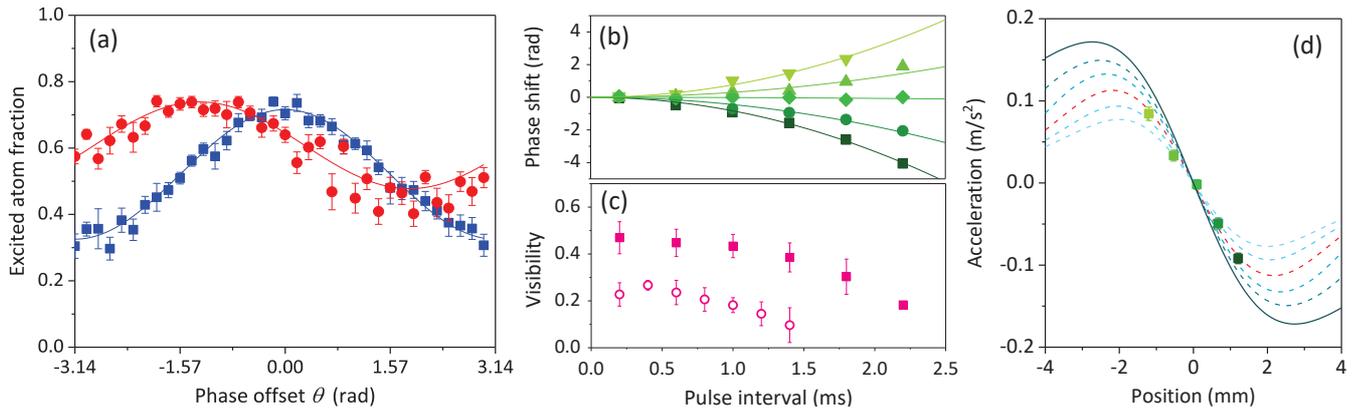}
\caption{
(a) Ramsey fringes measured for atoms at $x_0= 1.0$~mm with pulse intervals of $T=0.2$~ms (blue squares) and 1.4~ms (red circles). 
The phase of the first Ramsey pulse is offset by $\theta$ with respect to the following two pulses. 
Each point is integrated over 27 scans.
(b) Phase shift $\Delta \phi$ and (c) the visibility $A/B$ of the Ramsey fringes with respect to the pulse interval $T$. 
The depth of the colors corresponds to initial position of atoms $x_0=-1.20,\ -0.53,\ 0.09,\ 0.67,\ 1.21$~mm, from light green to dark green.
The phase shifts are fitted by a parabola (solid curves).
Empty circles show the visibility of MZI without velocity selection.
(d) Accelerations with respect to the atom position are indicated by filled squares with the corresponding depth of colors.
The curves show the accelerations calculated for different radial temperatures $T_{\rm r}$ of atoms. 
The measured acceleration agrees with the red-dashed curve corresponding to $T_{\rm r}=1.2~\mu$K.
}
\end{center}
\end{figure*}%

Figure~4(d) summarizes the atom position- dependent accelerations $a$. 
The solid line shows the acceleration of atoms without radial motion, i.e., $T_{\rm r}=0$, which gives the maximum acceleration with an oscillation frequency of $\nu_x=1.7$~Hz.
As the atoms are distributed radially owing to thermal motion, they experience less guide intensity and axial acceleration.
The dashed lines show position-averaged acceleration for the radial temperatures $T_{\rm r}=0.4,\ 0.8,\ 1.2,\ 1.6,$ and $2.0~\mu$K. 
The observed accelerations (filled squares) are well-explained with the red-dashed line with the radial temperature of $T_{\rm r}=1.2~\mu$K. 
This line represents the axial oscillation frequency of $\nu_x=1.4$~Hz, which is consistent with our observations in Fig.~3(b).

As an inertia sensing device, the major limitation in the free-space Gaussian beam guide comes from spatial variation of the guiding potential, which perturbs the external acceleration to be measured.
This extra acceleration of the guide is largely removed by using an optical guide formed inside a hollow-core photonic crystal fiber (HC-PCF), which allows the attenuation of the guide laser to be less than 1~dB/m \cite{Wan11} or allows an extinction coefficient of $\gamma=0.23~/{\rm m}$.
Assuming a guiding potential of $U_0=h\times 100$~kHz, which is sufficient to confine laser-cooled atoms, the additional acceleration is 
$a_0= -(1/m)[{\partial ({U_0}{e^{-\gamma x}})}/{\partial x}] \approx {\gamma {U_0}}/m \approx 10^{-4}~ {\rm {m/s^2}}$. 
This offset is canceled by alternately introducing the guide laser from opposite directions. 
The use of heavy atoms or a low-loss fiber reduces this correction and the relevant uncertainties.
This magic-guide interferometer can be best applied to an acceleration gradiometer~\cite{Sna98,Gra13,Hu17} that measures $\Delta a=a_2-a_1$ for atoms prepared at $x_2$ and $x_1$ to share the same clock laser as depicted in Fig.~1(b).
In this case, the clock-laser instability previously discussed is common-mode rejected by $\Delta\phi_2-\Delta\phi_1$, which allows approaching the quantum projection noise limited measurement, as demonstrated in the synchronous operation of clocks~\cite{Tak11,Ush15}.
Such AIs on the single-photon transition are also useful for a large-scale gradiometer , as proposed in Ref.~\cite{Gra13}.

In summary, we demonstrated a ``magic optical guide'' AI operating on the $^1\!S_0-{}^3\!P_0$ clock transition of $^{87}$Sr atoms.
To demonstrate the feasibility of the internal-state labeled MZI in the state-insensitive guide, the acceleration of atoms in the focused optical guide along the beam axis is measured.
To avoid the position-dependent acceleration that is intrinsic to the free space guide, we proposed a magic guide MZI and gradiometer inside the HC-PCF, where the internal-state labeling plays an important role in miniaturizing the device. 
Such a device is also useful for investigating tens of $\mu$m scale gravitational interaction. 
All the components of the magic guide interferometer are the same as those used for optical lattice clocks, such as fiber clocks~\cite{Oka14} and synchronously operated clocks~\cite{Tak11}. 
Turning off the lattice allows the clocks to be inertial sensors.

\subsection*{Acknowledgments}

We thank Y. Komine for his support in the early stage of the experiment and A. Hinton for carefully reading of the manuscript.
This work is supported by JST ERATO Grant Number JPMJER1002 (Japan), by JSPS Grant-in-Aid for Specially Promoted Research Grant Number JP16H06284, and by the Photon Frontier Network Program of the Ministry of Education, Culture, Sports, Science and Technology, Japan.


\end{document}